\documentclass[11pt,preprint]{aastex}
\usepackage{epsfig}
\received{} \shortauthors {Shaw et al.} \shorttitle {Confirmation of New LMC Planetary Nebulae}

\begin{document}

\title {Confirmation of New Planetary Nebulae in the Large Magellanic Cloud\footnote
{Based on observations with the NASA/ESA Hubble Space Telescope,
obtained at the Space Telescope Science Institute, which is operated
by the Association of Universities for Research in Astronomy, Inc., under
NASA contract NAS 5-26555.}
}

\author {Richard A.~Shaw}
\affil {National Optical Astronomy Observatory, Tucson, AZ 85719}
\email {shaw@noao.edu}
\and
\author {Warren A. Reid\altaffilmark{2}, Quentin A. Parker\altaffilmark{2}}
\affil {Department of Physics, Macquarie University, Sydney, NSW 2109, Australia}
\email {warren@ics.mq.edu.au, qap@ics.mq.edu.au}

\altaffiltext{2}{Anglo-Australian Observatory, PO Box 296, Epping, NSW 1710, Australia}


\begin{abstract}

We present {\it Hubble Space Telescope} ({\it HST}) images of new planetary nebulae (PNe) 
that were discovered in the Reid-Parker AAO/UKST H$\alpha$ survey of the Large Magellanic 
Cloud. These serendipitous 
observations from various {\it HST} programs yield independent confirmations of 6 PNe; one 
other detected nebula may also be a PN, and one appears to be a region of diffuse emission. 
The high resolution {\it HST} archival images enable us to determine the physical sizes, the 
nebular morphology, and related features of these new PNe in detail for the first time. In a few 
cases we were also able to identify the central star, which was impossible with the lower 
resolution, wide-field discovery data. The confirmation of faint, extended halos surrounding 
many PNe in the RP catalog must await the acquisition of new deep, high-resolution, 
narrow-band imagery. 

\end{abstract}


\section {Introduction}

The study of planetary nebulae (PNe) in the Magellanic Clouds spans half a century, but it 
has received renewed attention in the past half decade. New discovery surveys, detailed 
spectroscopic studies, and a greater understanding of the Magellanic Clouds themselves 
has greatly increased our understanding of PNe in these systems. 
Studying PNe in the Magellanic Clouds offers significant advantages compared to studies 
of the Galactic PN populations \citep[see e.g.,][]{Jacoby03, Shaw06}. Chief among them is 
their location in systems at known distance so that large numbers of PNe can be studied in 
detail, and important physical parameters such as sizes and luminosities can be determined 
with high accuracy. The average foreground extinction to both the Large Magellanic Cloud 
(LMC) and Small Magellanic Cloud (SMC) is low so 
that large, complete, flux-limited samples may be obtained without the severe selection 
biases that apply in the Galaxy. Finally both systems are among the most massive in the 
Local Group, so that the PN populations total hundreds of objects which enables the study 
of statistical properties of various sub-samples. Magellanic Cloud PNe have thus been very 
successfully employed over the past few decades to investigate a number of very important 
astrophysical questions such as the form and origin of the PN luminosity function, the effects 
of metallicity and environment on PN properties and the late stages of stellar evolution, and 
as kinematical probes of the galaxies themselves \citep[e.g.,][]{RP06b}. More recently, high 
resolution imaging and spectroscopy have been employed to study the coevolotion of 
individual nebulae \citep{Shaw_etal01, Stang_etal02, Stang_etal03, Shaw_etal06} and 
their central stars \citep{Villa03, Villa04, Villa07}. 

Observing PNe in the Magellanic Clouds is not without its challenges, however. The fields are 
extremely crowded, the individual nebulae cannot be resolved in useful detail except with 
{\it HST} or advanced optical systems on large, ground-based telescopes, and the LMC in 
particular is permeated with very bright, spatially complex nebular emission from a variety of 
sources including \ion{H}{2} regions, supernova remnants, and wind-blown nebulae around 
O-stars. In addition, the Magellanic Clouds extend over a wide area of sky. These issues 
complicate the task of conducting uniform, deep surveys for PNe over the full spatial extent 
of these galaxies. The heterogeneity of prior surveys in the spatial coverage, depth, discovery 
technique, and the lack of complete follow-up spectroscopy for verifying the classification 
have been the principal barriers to realizing the full potential of Magellanic Cloud PN studies 
until recently. 

Recent surveys with wider areal coverage, uniform depth, and greatly increased sensitivity 
have been published by \citet{JD02} for the SMC and by \citet{RP06a, RP06b} for the LMC; 
another in the LMC is underway \citep{Smith99}. See \cite{Jacoby03} and \cite{Parker06} 
for recent reviews of these and other surveys. We focus in this paper on PNe in the LMC, 
the known population for which has increased greatly with the publication of an extensive 
catalog by \citet[][hereafter referred to as the RP catalog]{RP06a, RP06b}. The RP catalog 
is based upon both a deep photographic survey in H$\alpha$ and in red continuum of the 
central 25~square degrees of the LMC, and upon follow-up confirmatory spectroscopy. 
In all, 460 new objects were classified as PNe with varying degrees of confidence: a  
thorough analysis of optical emission line ratios, nebular morphological features, prominence 
of the continuum, and the inferred degree of contamination from background sources, resulted 
in 292 objects being classified by \citet{RP06a, RP06b} as ``true" PNe, 54 as ``likely," and 
114 as ``possible." The lack of complete certainty in the PN classification is in part due to 
the similarity of emission line ratios between very low excitation (VLE) PNe and other kinds 
of objects, and also to strong, diffuse \ion{H}{2} emission or strong contaminating continuum 
emission in the immediate vicinity of some objects. 

The thrust of this study was to search the {\it HST} archive for images that might contain any 
of the new RP catalog objects, and to use any detections to illuminate a) the veracity of the 
PN designations, particularly for cases were the classifications in the RP catalog were 
uncertain, b) the nebular dimensions and morphological features, c) the location and 
characteristics of the central stars, and d) strategies for follow-up observations of the RP sample. 
In \S2, we describe the RP catalog of LMC PNe and the survey upon which it is based, and 
then describe the observational data extracted from the {\it HST} archive that matches the 
new sources in the RP catalog. The individual objects are described in \S3, along with an 
analysis of non-detections. In \S4 we discuss the implications of these data for future 
observing programs.

\section {Observational Data}

The surveys upon which the RP catalog of LMC PNe is based, the observing strategies, 
the discovery technique, the follow-on spectroscopy, and the classification schema are all 
described by \citep{RP06a,RP06b}. We summarize in the next subsection some of the key 
features of the RP survey and discovery images in enough detail to explain the comparison 
to the serendipitous {\it HST} images. Following that we describe the attributes of the 
{\it HST} observations, the reduction procedures, and our analysis of the data. 

\subsection {AAO/UKST H$\alpha$ Deep Stack Images}

A photographic mini-survey of 40 fields in and around the Magellanic Clouds was undertaken 
as part of the Anglo-Australian (AAO)/UKST H$\alpha$ 
survey of the Southern Galactic Plane and Magellanic Clouds \citep{Parker05}. In addition, 
a separate, deep AAO/UKST photographic survey of the central 25~square degrees of the 
central bar of the LMC was also undertaken by Q.~Parker and D.~Morgan between 1998 and 
2000. Twelve well-matched, 2 hour H$\alpha$ exposures and six 15 min equivalent short 
red (SR) broad-band exposures on this field were taken for the purpose. The images were 
scanned by the ``SuperCOSMOS'' measuring machine at the Royal Observatory Edinburgh 
\citep{Hambly_etal01}, and the resulting digitized images were placed on a common world 
coordinate system and co-added. The photometric depth is $R\sim$21.5 mag for the SR 
images and $R_{equiv}\sim$22 mag for H$\alpha$ 
(or in units of flux density, $4.5\times10^{-17}$ erg~cm$^{-2}$~s$^{-1}$~\AA$^{-1}$). 
The pixel size of the scanned images is 10$\mu$m (0\farcs67) which samples the image 
sufficiently well to resolve compact nebulae larger than 3\farcs5 in diameter. Spurious 
detections such dust particles were effectively removed through the median-stacking of the 
individual exposures taken over a 3 year period while the influence of variable stars was 
considerably abated. This new H$\alpha$ map led directly to the discovery of a significant 
new population of LMC emission sources, including the many new PNe that comprise the
Reid-Parker catalog of LMC PNe.  

Candidate emission sources were identified using an adaptation of a technique available 
within the KARMA collaboration \citep{Gooch96} which is fully described by \cite{RP06a}. 
The stacked SR and H$\alpha$ images were each assigned a specific color: red for SR and blue 
for H$\alpha$. The colored images were then merged, resulting in a pink color for continuum 
sources, whereas compact or point source emitters develop a blue aura around a pink core 
whose relative size depends on the intensity of the emission. Extended nebulae, including 
\ion{H}{2} regions, supernova remnants, super-bubbles and resolved, faint PNe can appear 
solely blue on an otherwise dark background. Many of the PNe are only visible as blue 
features in the false-color image. Most PNe central stars however are too faint, and the 
resolution too poor in the ground based UKST images, to be detected. 
A variety of telescopes and instruments were employed for spectroscopic follow-up including 
FLAMES on the Very Large Telescope (VLT) UT2, the 1.9-m telescope at the South African 
Astronomical Observatory, the 2.3-m telescope at Siding Spring Observatory, and 6dF on the 
UKST.  However, the vast majority were confirmed using 2dF on the Anglo Australian 
Telescope (AAT) in 2004 December \citep[see][]{RP06b}. 

\subsection {AAO 2dF Fiber Spectra}

Confirmatory spectra were obtained for nearly every object in the RP catalog \citep{Reid06}. 
All of the spectra for the objects presented here were obtained with the AAO 2dF 
spectrograph on the AAT, which is capable of obtaining more than 400 simultaneous 
spectra, using a 300 line mm$^{-1}$ grating to cover the spectral range 3650--8050\AA\AA\
with a dispersion of 4.3 \AA\ pixel$^{-1}$. 
The spectra were corrected for detector bias, extracted, and sky-subtracted. The 
wavelength calibration was performed with the aid of comparison arcs taken during the 
course of the observations, but no flux calibration was performed. See \citet{RP06b} and 
\citet{Reid06} for details of the reductions. The spectrograms are shown in 
Figures~\ref{Spec_1} and \ref{Spec_2}; the flux scale is in instrumental units. 

\subsection {HST Images}

Our search of the {\it HST} archive (conducted in 2006 June) identified nearly 200 images that 
were coincident with 19 of the newly discovered PNe candidates in the RP catalog, or 
about 4.0\% of the sample of new PNe. Such a high number of serendipitous observations is 
due in large part to the popularity of the Magellanic Clouds for various {\it HST} research 
programs, particularly for stellar populations. Many additional exposures were obtained 
through pure parallel programs. Although the majority of the objects were observed with 
more than one filter, more often than not these were broad-band filters which were not well 
suited for detecting emission line nebulae, in that they do not include the strongest nebular 
emission lines within the bandpass. However, when available these filters were useful for 
identifying extremely blue stars in the vicinity, which aided the identification of the likely 
source of the nebular ionization. Fortunately, some of the archival images that match RP 
catalog positions were taken through more suitable filters. 
The WFPC2/F300W bandpass includes nebular [\ion{O}{2}] $\lambda3727$ and 
[\ion{Ne}{3}] $\lambda3889$ emission when present, but the system efficiency is quite low. 
The F555W filter includes H$\beta$ and the strong [\ion{O}{3}] $\lambda\lambda4959+5007$ 
emission lines, plus H$\alpha$ + [\ion{N}{2}] $\lambda\lambda6548+6583$ at lower efficiency. 
F606W, which has the highest overall system efficiency, includes the [\ion{O}{3}], H$\alpha$, 
and [\ion{N}{2}] emissions mentioned above. The clearest detections were made with the 
F656N filter (which includes some [\ion{N}{2}] emission in the wing), but this filter is not as 
often used. 

The {\it HST} images presented here were obtained using the Wide-Field Planetary 
Camera 2 (WFPC2) or the Advanced Camera for Surveys (ACS). These images were 
calibrated with the standard reduction pipelines for the instruments used to obtain them. 
The pipelines perform similar processing: bias correction, dark scaling and subtraction, and 
flat-fielding. The ACS pipeline also performs a geometric rectification. For many of the 
objects, multiple exposures were obtained at or near the same location with the same 
instrument and filter combination. Typically this is done to facilitate the identification and 
rejection of cosmic rays, to mitigate the effects of detector defects such as charge traps or 
high dark current, or (in the case of WFPC2) to improve the spatial sampling by dithering 
the image. These images were combined when possible. For WFPC2 many pre-combined 
images are available from the  Multi-mission Archive at STScI (MAST) as high level science 
products (HLSP). \citet{Wada_etal06} describe the details of this processing, as well as the 
quality of the photometry of the resulting images. For the present purpose we are concerned 
primarily with determining precise coordinates, sizes, and morphologies of the new RP 
nebulae, so that the accuracy of the reductions is not critical. However we did combine images 
taken with the same filter, when possible, to eliminate cosmic rays and to improve the 
signal-to-noise ratio for presentation.
The {\it HST} observing log for all nebulae is presented in Table~\ref{ObsLog}. The first 
column lists the RP designation, followed by the confidence with which \citet{RP06b} could 
establish the target as a genuine PN based on their imaging and confirmatory spectroscopy, 
followed by their estimate of the diameter of the nebular shell. This information is followed in 
the next two columns by the {\it HST} observing configuration, followed 
by the exposure time and whether extended emission was detected. The last column lists the 
dataset identifiers in the MAST archive, which corresponds to high-level science products (i.e., 
combined images as generated through, e.g., the WFPC2 Archival Pure Parallels Project 
pipeline) where available. 

\section {Dimensions and Morphology for the Matched Sources}

The {\it HST} and matching RP catalog images for the detected nebulae are presented 
in Figures~\ref{Neb_1} through \ref{Neb_4}, where for each object we show both the UKST 
discovery image and the {\it HST} image on the same spatial scale. All of the images
are rendered in false-color, usually with a square-root intensity scale, in order to bring out 
the often faint structural features of the nebulae. In spite of the generally low signal-to-noise 
(S/N) ratio in the {\it HST} images (which, after all, were not obtained for this purpose), we 
were able to confirm several of the objects as genuine PNe. Furthermore the 0\farcs1 
resolution of the {\it HST} images permits a solid morphological classification to be assigned, 
which is not possible from the \citet{RP06b} data. The morphological classifications would 
ordinarily be based on the appearance of the nebula in the monochromatic light of either 
[\ion{O}{3}] $\lambda5007$ or H$\alpha$ (possibly including [\ion{N}{2}] 
$\lambda\lambda6548+6583$). Although WFPC2 monochromatic emission-line images are 
only available for three of the objects, \cite{Shaw_etal06} point out that the principal, 
larger scale morphological features can generally be inferred from other emission lines with 
some confidence. We used the same classification scheme as those of \citet{Shaw_etal01,
Shaw_etal06} and \cite{Stang_etal02, Stang_etal03}, where the morphological types of 
relevance here are: round (R), elliptical (E), and bipolar (B). Other important structural 
features, such as attached shells, were noted as well. 

Table~\ref{Morph} gives detailed information for the PNe presented here. The equatorial 
J2000 sky coordinates, as measured from the {\it HST} images, are given in columns (2) 
and (3). These coordinates correspond to either the geometric centroid of the nebula or, in 
three cases, the newly identified central stars. They agree to within 1\arcsec\ with 
the (likely more accurate) RP positions, which is consistent with the accuracy of the 
{\it HST} guide star frame of reference \citep[GSC 1.1,][]{Morrison_etal01}. Nebular 
diameters along the major and minor axes, given in column (4), were usually measured with 
respect to the 20\% intensity contour of the outermost structure that was visible in the 
{\it HST} image. The dimensions given in Table~\ref{Morph} are generally smaller than 
those cited by \citet{RP06b}. The reasons for the discrepant sizes are varied: 
a) Nebulae smaller than 3\farcs4 were not resolved in the RP data; 
b) one or more nebulae (probably including RP1550) are surrounded by faint, outer shells 
or halos that are not detected in these HST exposures; 
c) differing measurement techniques were employed: here we prefer the 20\% intensity 
contour, in part because of the limited S/N ratio of the {\it HST} data and the impracticality 
of measuring a photometric radius, whereas \citet{RP06b} measured the full width at zero 
intensity in monochromatic light; 
d) nebulae embedded in a bright continuum were problematic to measure in the RP 
images owing to the response of the photographic media, as explained by \citet{RP06b}. 
Our assessment of whether the object is a genuine PN is given in column (5),
and the nebular morphological classification is given in column (6). Column (7)
contains specific notes about each object, including whether a central star was
identified.

\subsection {Detected Objects}

As discussed in the previous section, extended emission was detected in the {\it HST} 
archival images for eight RP catalog objects. We describe here the morphological details 
and noteworthy spectral features for each of the nebulae listed in Table~\ref{Morph}. 

\begin {figure}
\plotone {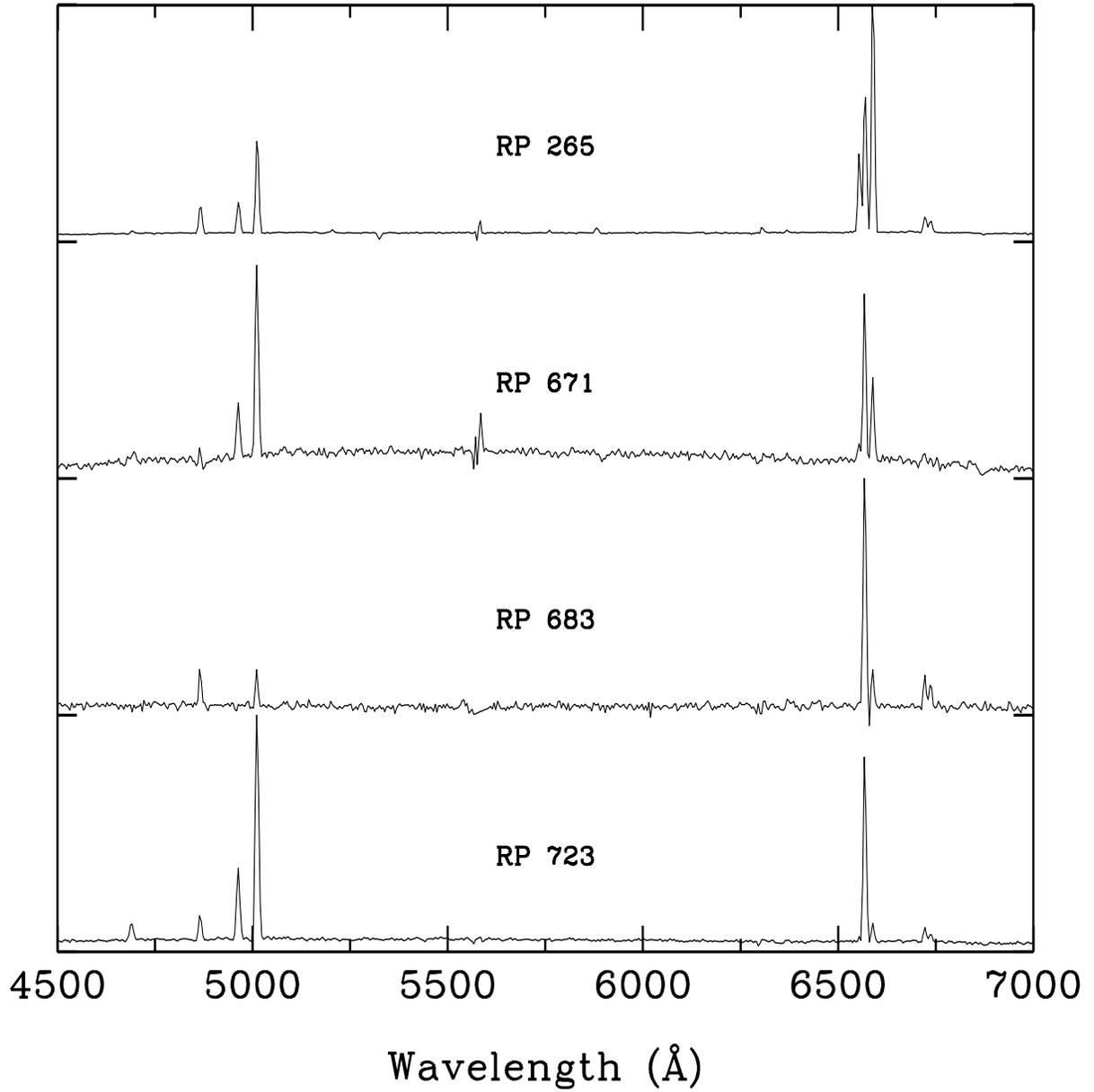}
\figcaption[f1.eps]
{Spectrograms of RP265, RP671, RP683, and RP723 (top to bottom), displaced 
vertically for clarity. Brightness is in units of normalized instrumental counts. 
\label{Spec_1}}
\end {figure}

{\it LMC--RP~265.---}The {\it HST} image for this object was presented by 
\citet{Shaw_etal06}. Although it is extremely faint, it is probably a bipolar. The 
spectrum in Figure~\ref{Spec_1} indicates a moderate excitation, owing to the modest 
[\ion{O}{3}]/H$\beta$ ratio and weak \ion{He}{2} $\lambda$4686 emission. The very 
strong [\ion{N}{2}]/H$\alpha$ ratio is consistent with the strong nebular bi-polarity 
\citep{Shaw_etal06}. 

\begin{figure}
\plotone {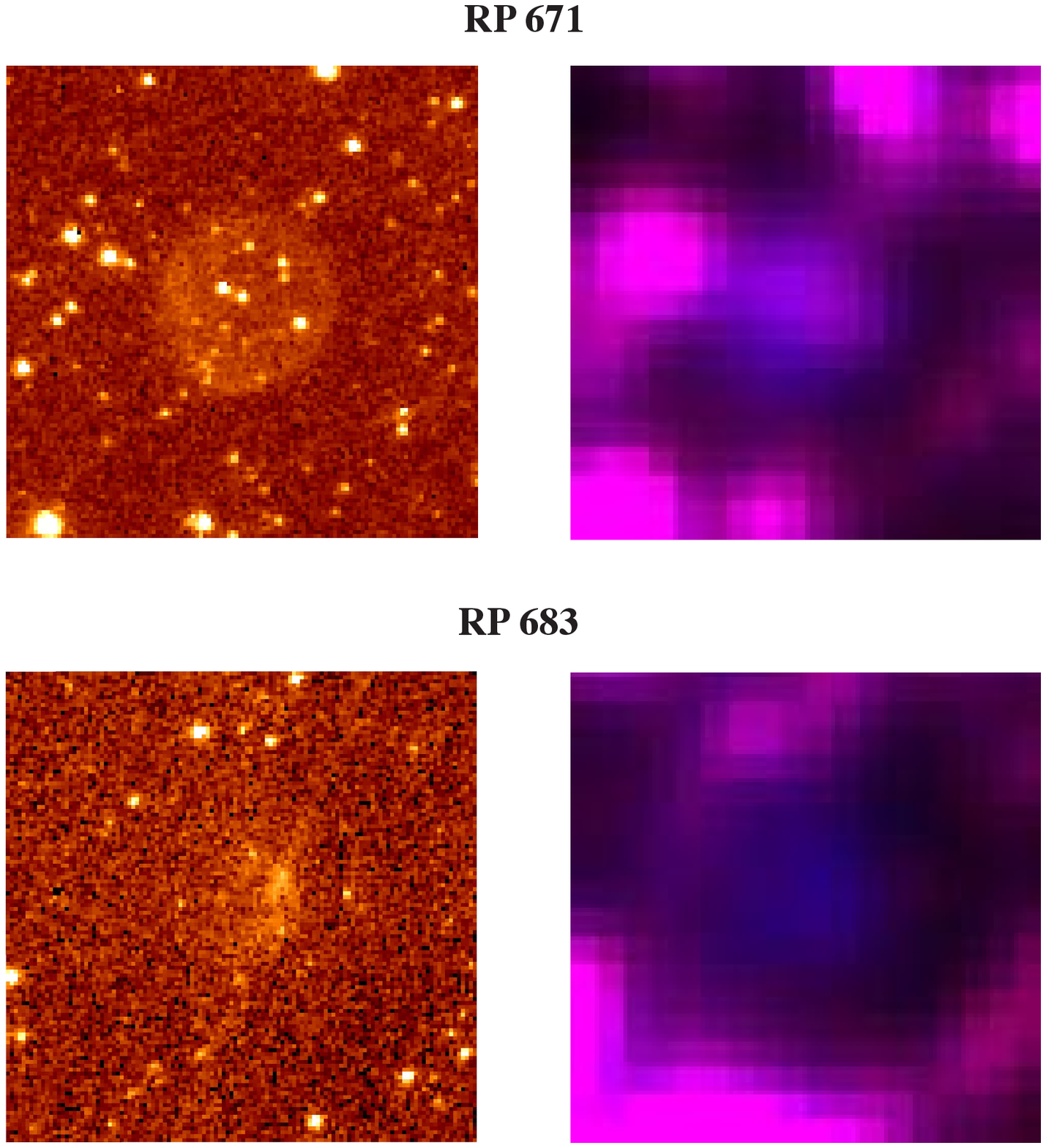}
\figcaption[f3.eps]
{False-color images of RP nebulae from {\it HST} ({\it left}) with a square-root intensity 
stretch, and matching, false-color UKST merged H$\alpha$ + SR images ({\it right}), 
from which the objects were discovered (see text). 
The nebulae are RP671 ({\it top}) and RP683 ({\it bottom}); both {\it HST} images are with 
WFPC2/F656N. 
All images are 12\arcsec\ on a side, with north upwards and east to the left. 
\label{Neb_1}}
\end {figure}

\begin {figure}
\plotone {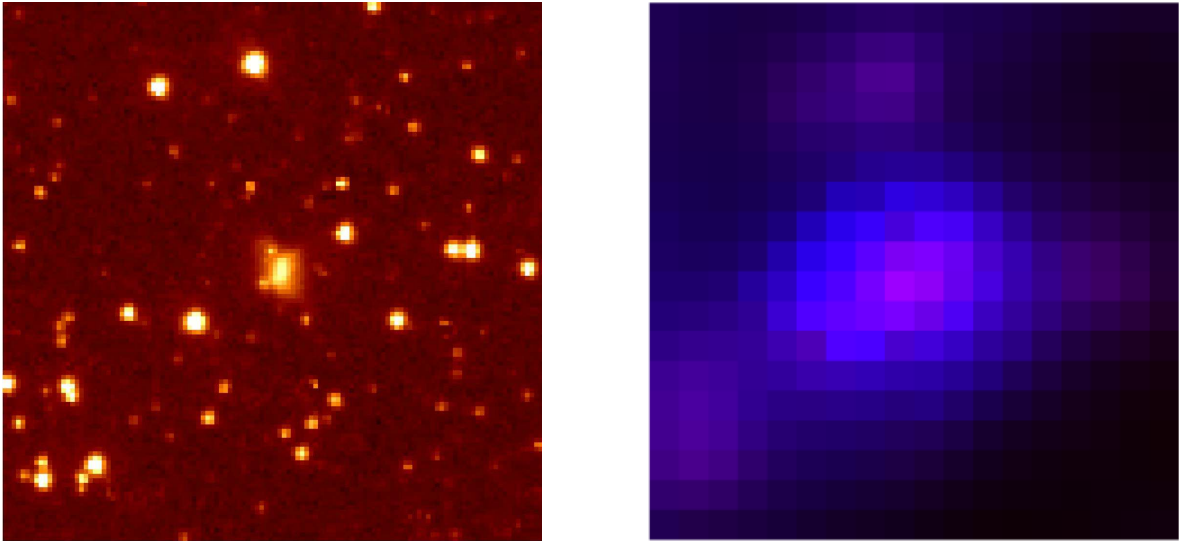}
\figcaption[f6.eps]
{Same as Fig.~\ref{Neb_1} for RP1550 with WFPC2/F555W. 
Cosmic ray removal from WFPC2 single frame may be incomplete. 
Image scale, orientation, and intensity stretch are as in Fig.~\ref{Neb_1}. 
\label{Neb_4}}
\end {figure}

{\it LMC--RP~671.---}This object is confirmed as an extremely low surface brightness PN.
It is very apparent in the combined 5000~s exposure with the WFPC2/F656N filter, but it is
barely detected in F450W and F606W images. This nebula is large, nearly perfectly 
round, and is similar in appearance to many Galactic Abell PNe (see Figure~\ref{Neb_1}). 
The central star is near the geometric center of the nebula, and is easily identified
through its $U-B$ color. The spectrum in Figure~\ref{Spec_1} indicates a fairly high 
excitation, with a large [\ion{O}{3}]/H$\beta$ ratio and relatively strong \ion{He}{2} 
$\lambda$4686 emission, which is consistent with an advanced evolutionary state for 
this star + nebula. 

{\it LMC--RP~683.---}This apparently diffuse nebula may be ionized by one or more of a 
number of very blue stars (judging from the $U-R$ colors) that lie within $\sim$10\arcsec, 
one of which lies within (or is projected onto) the nebulosity. RP categorize this object as 
only a ``possible" PN, and also a very low excitation (VLE) object given that the intensity 
of H$\beta$ is similar to that of [\ion{O}{3}] $\lambda5007$ (see Figure~\ref{Spec_1}). Given 
the morphology evident in the {\it HST} imagery (Figure~\ref{Neb_1}) it is unlikely that this 
object is a genuine PN. 

\begin {figure}
\plotone {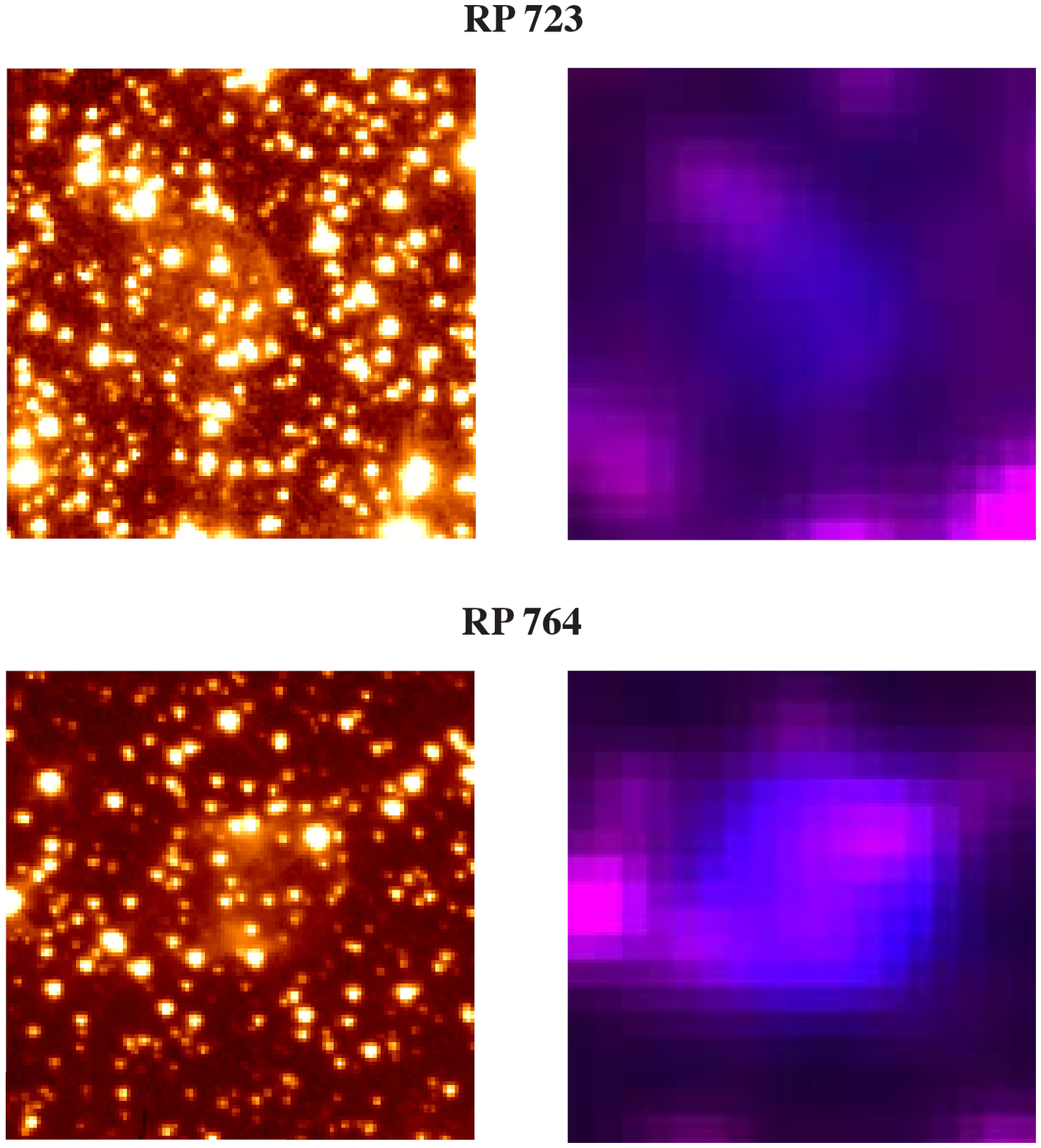}
\figcaption[f4.eps]
{Same as Fig.~\ref{Neb_1} for RP723 with WFPC2/F555W, and RP764 with 
WFPC2/F606W. 
Only the brightest parts of the PN shell in RP764 are apparent in the {\it HST} image, 
although there is a suggestion of emission in the interior (east and west sides). 
The full extent of the shell is clearly visible in the UKST image. 
Image scales, orientations, and intensity stretches are as in Fig.~\ref{Neb_1}. 
\label{Neb_2}}
\end {figure}

{\it LMC--RP~723.---}This object is an extremely low surface brightness PN. It is very faint 
in the moderate-length exposure with the WFPC2/F555W filter. The nebula is round, 
although the western limb is brighter and has a more sharply defined edge (see 
Figure~\ref{Neb_2}). The likely central star appears to be near the geometric center of the 
nebula, and is tentatively identified through its $B-V$ color. The spectrum in 
Figure~\ref{Spec_1} indicates a fairly high excitation, with a large [\ion{O}{3}]/H$\beta$ 
ratio and relatively strong \ion{He}{2} $\lambda$4686 emission. 

\begin {figure}
\plotone {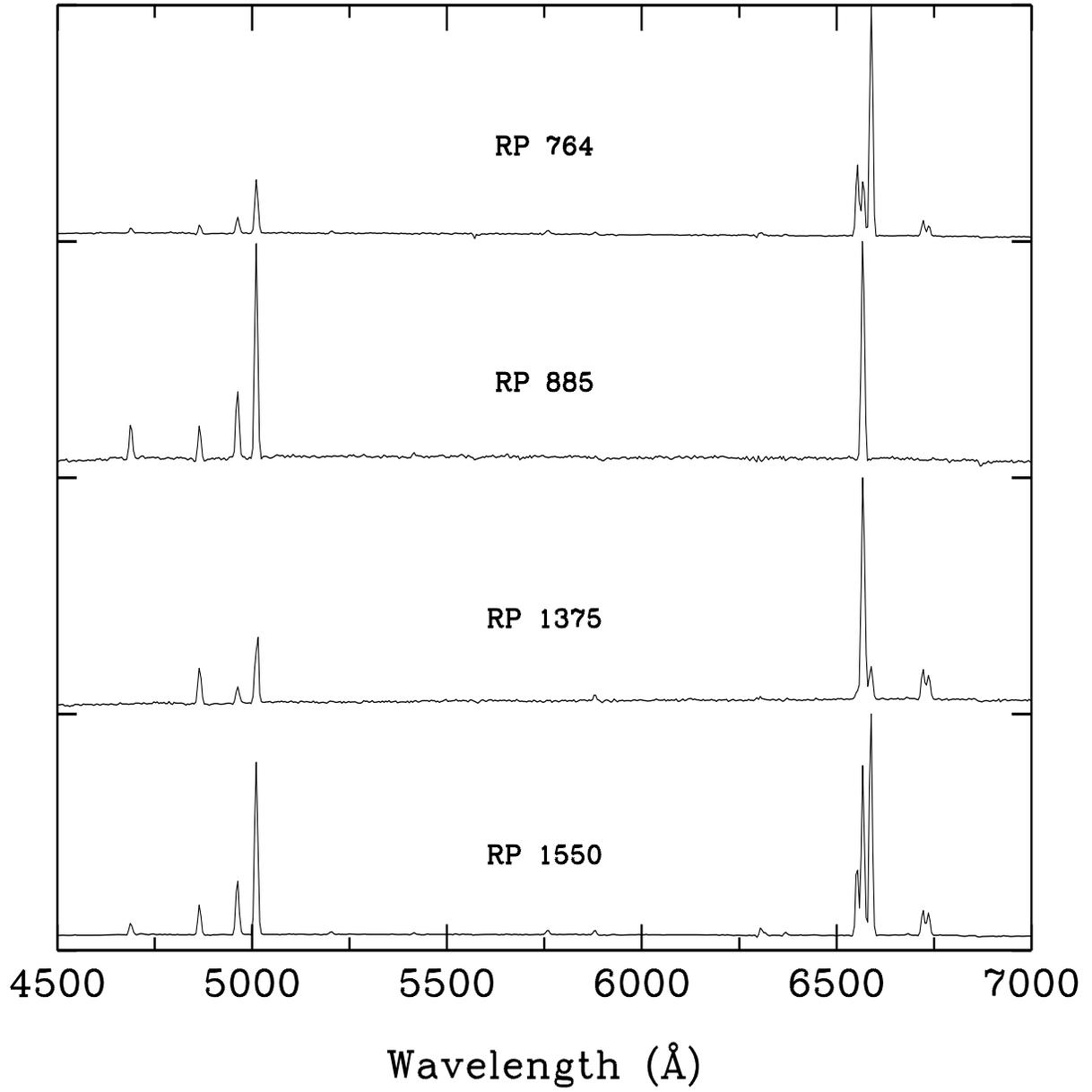}
\figcaption[f2.eps]
{Same as Fig.~\ref{Spec_1} for RP764, RP885, RP1375, and RP1550 (top to bottom). 
\label{Spec_2}}
\end {figure}

{\it LMC--RP~764.---}This object is clearly a bipolar PN and is similar in appearance to the 
Dumbbell nebula (see Figure~\ref{Neb_2}). The identification of the central star is puzzling: 
based on the F330W image, a star at the northwest edge of the nebula is by far the bluest in the 
image, and is the best candidate. The spectrum in Figure~\ref{Spec_2} indicates a moderate 
level of excitation, with a moderate [\ion{O}{3}]/H$\beta$ ratio and finite \ion{He}{2} 
$\lambda$4686 emission. The very strong [\ion{N}{2}]/H$\alpha$ ratio (more than 4:1) is 
consistent with the strong nebular bi-polarity \citep{Shaw_etal06}. 

\begin {figure}
\plotone {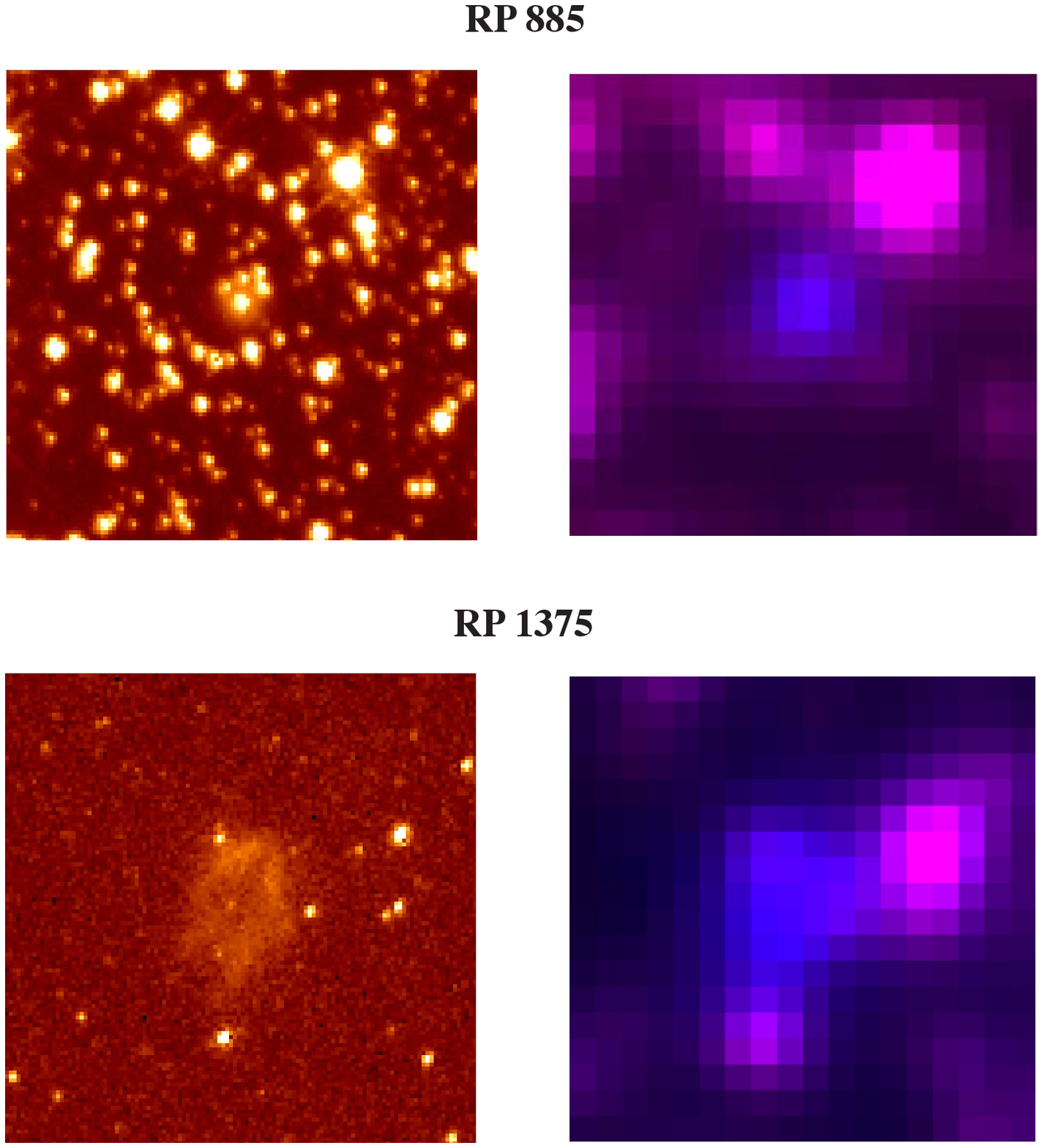}
\figcaption[f5.eps]
{Same as Fig.~\ref{Neb_1} for RP885 with WFPC2/F555W, and RP1375 with 
WFPC2/F656N. 
Image scales, orientations, and intensity stretches are as in Fig.~\ref{Neb_1}. 
\label{Neb_3}}
\end {figure}

{\it LMC--RP~885.---}This object is confirmed as a round, faint PN. It has a main shell of
$\sim0.27$ pc in diameter, and has an attached shell that is twice as large but fainter by
a factor of at least 3 (see Figure~\ref{Neb_3}). The central star is relatively bright. The 
spectrum in Figure~\ref{Spec_2} indicates a very high excitation, with a large 
[\ion{O}{3}]/H$\beta$ ratio and very strong \ion{He}{2} $\lambda$4686 emission. The 
absence of [\ion{N}{2}] emission suggests that N is very under-abundant, which is 
consistent with the nebular morphology. 

{\it LMC--RP~1375.---}The extremely low surface brightness of this object makes a 
definitive classification as a PN difficult, in spite of the lengthy exposure time in the 
WFPC2/F656N filter, but it could be a genuine PN. The morphological classification is very 
uncertain: the overall shape is elliptical, but it has internal structure that suggests a bipolar 
core (see Figure~\ref{Neb_3}). The central star might be the star near the geometrical center 
of the nebula, but no other passbands are available for confirmation. The spectrum in 
Figure~\ref{Spec_2} show a low level of excitation, with a low [\ion{O}{3}]/H$\beta$ ratio, 
no \ion{He}{2} $\lambda$4686 emission, but weak \ion{He}{1} $\lambda$5876 emission. 

{\it LMC--RP~1550.---}Despite the relatively small angular size this object is confirmed 
as a bipolar PN, even with the short exposure time in the WFPC2/F555W filter (see 
Figure~\ref{Neb_4}). No central star is evident in the image. The spectrum in Figure~\ref{Spec_2} 
indicates a moderately high level of excitation, with a high [\ion{O}{3}]/H$\beta$ ratio and 
moderate \ion{He}{2} $\lambda$4686 emission. The strong [\ion{N}{2}]/H$\alpha$ ratio 
is consistent with the strong nebular bi-polarity \citep{Shaw_etal06}. 

\subsection {Non-Detections}

Of the 19 RP sources that were imaged with {\it HST}, extended emission was not detected 
around eleven of the targets. It is useful to consider whether the non-detections are 
significant, particularly for RP sources classified as ``possible" PNe. In most cases the 
non-detections can be attributed either to very brief exposure times of less than a minute 
(3 objects: RP307, RP1759, RP2180), or to the use of filters that do not include strong nebular 
emission lines (4 objects: RP232, RP241, RP1443, RP1580). In three other cases, it is 
possible that the nebular emission is extended, but with a surface brightness that is lower 
than the detection threshold in the {\it HST} image. To explore this possibility, we make use 
of the nebular diameters from \citet{RP06b}, the empirical relation of the decline of average 
[\ion{O}{3}] surface brightness with radius from \citet{Shaw_etal06}, and assuming the 
[\ion{O}{3}] $\lambda5007$ flux is comparable in intensity to H$\alpha$. The exposure time 
calculator for ACS (scaling by the relative efficiency for WFPC2 images where necessary) 
provides some indication of whether extended emission in the remaining objects might have 
been detectable, given the above assumptions. In one case, RP218, the average surface 
brightness would have been far below the detection threshold for WFPC2/F606W. But for two 
objects we might have expected to detect extended emission: RP203 with a signal-to-noise 
(S/N) ratio of about 2, and RP268 with a S/N ratio of about 4. Given the dispersion in the surface 
brightness relation and in the other assumptions, it is easily possible that the emission from 
RP203 and RP268 is extended, but with an emissivity that lies just below the detection 
threshold. For RP505 the extremely weak confirmation spectrum lead \citet{RP06b} to 
classify it as only a ``Possible" PN that is not resolved in their images. It may be that this 
object is a faint, diffuse nebula like RP683. 

\section {Discussion and Summary}

Our search of the {\it HST} archive for serendipitous observations of new LMC PN 
candidates in the RP emission-line survey has resulted in the independent confirmation 
of 8 extended emission objects, 6 of which can be confirmed as genuine 
PNe including one that was independently confirmed as a PN by \citet{Shaw_etal06}. 
There is also one source, RP1375, where \citet{RP06b} classified it as a ``possible" 
PN and the morphology from the {\it HST} image is uncertain. Finally, RP683 is a region 
of diffuse emission that may be an \ion{H}{2} region; but which was only classed as a 
possible VLE PN by \citet{RP06b}. The central stars can be readily identified in four of 
the PNe, and possibly in a fifth. Perhaps as importantly, it is very clear from the {\it HST} 
images the degree of crowding of stars in the vicinity of each object, which will guide 
the strategy of follow-up studies. 
We found {\it HST} images of fields surrounding an additional 11 RP objects, but no 
extended emission was detected at the published coordinates. A close examination 
shows that the non-detections can always be explained by a combination of inappropriate 
filters and exposures of insufficient depth to reach the anticipated nebular surface 
brightness. 

All of the RP nebulae identified here are faint and most have very low average surface 
brightness, which are characteristics common to many of the new PNe in the RP catalog. 
While these serendipitous, high-resolution images provide a useful validation of the PN 
morphological classifications, they did not prove to be very useful for detecting or 
validating the faint, outer shells or halos that \citet{RP06b} found for nearly 60\% of all PNe in 
their catalog. Even in the extensive, targeted surveys by \citet{Shaw_etal01, Shaw_etal06} 
and \citet{Stang_etal02, Stang_etal03} faint halos were seldom detected. Very much 
deeper exposures with $\sim$0\farcs1 resolution will be required to study and understand 
these features in any detail. Still, the images in this study are another useful step in 
our understanding of the nature and evolution of planetary nebulae in the LMC. 

\acknowledgements

Support for this work was provided by NASA through grants GO-09077 and
GO-10251 from Space Telescope Science Institute, which is operated by the
Association of Universities for Research in Astronomy, Incorporated, under NASA 
contract NAS5--26555. RP thanks Macquarie University for a RAACE Ph.D. 
scholarship. 

\clearpage

%
%

\clearpage

%
%
\begin{deluxetable}{llcllrll}
\tabletypesize{\scriptsize}
\tablecolumns{8}
\tablewidth{0pt}
\tablecaption {Log of {\it HST} Observations \label{ObsLog}}
\tablehead {
\colhead {} & \colhead {RP} & \colhead {RP Diam.} & \colhead {Instrument/} & \colhead {} & 
\colhead {T$_{Exp}$} & \colhead {Nebula} & \colhead {} \\
\colhead {Nebula} & \colhead {Certainty} & \colhead {(arcsec)} & \colhead {Aperture} & 
\colhead {Filter} & \colhead {(s)} & \colhead {Detected} & \colhead {Dataset}
}
\startdata
LMC--RP~203 & True & 9.0 & ACS/WFC1& F555W &  250 & No & J8NE75IDQ \\*
   & & & & F814W & 170 & No & J8NE75J2Q \\
LMC--RP~218 & Possible & 14.0 & WFPC2/WFALL & F606W &  2x700 & No & U8IXN301M \\
LMC--RP~232 & Likely & 6.0 & WFPC2/WFALL & F450W & 240 & No &
U4WOEM02B\tablenotemark{a} \\*
   &  &  &  & F450W & 440 & No & U4WOEM04B\tablenotemark{a} \\*
   &  &  &  & F814W & 600 & No & U4WOEM06B\tablenotemark{a} \\*
   &  &  &  & F300W & 640 & No & U4WOEM08B\tablenotemark{a} \\
LMC--RP~241 & Possible & 6.0 & WFPC2/WFALL& F300W & 3660 & No & U2OU7N01b\tablenotemark{a} \\
LMC--RP~265\tablenotemark{b} & True & $5.4\times4.5$ & WFPC2/WFALL & F606W & 240 & Yes & U4WOEM02B\tablenotemark{a} \\*
   &  &  &  & F450W & 340 & No & U4WOE704B\tablenotemark{a} \\*
   &  &  &  & F814W & 500 & No & U4WOE706B\tablenotemark{a} \\*
   &  &  &  & F300W & 440 & No & U4WOE708B\tablenotemark{a} \\
LMC--RP~268 & Possible & 4.0 & WFPC2/WFALL& F606W &  2x300 & No & U4K2CQ01R \\
LMC--RP~307 & Possible & 4.0 & ACS/HRC & F220W &  20 & No & J8QZ02011 \\*
   &  &  &  & F435W & 10 & No & J8QZ02021 \\
LMC--RP~505 & Possible & 2.3 & WFPC2/WFALL& F606W &  240 & No & U4WOE202B\tablenotemark{a} \\
LMC--RP~671\tablenotemark{b} & True & 5.4 & WFPC2/WFALL & F606W & 1600 & Marginal & U4WODE04B\tablenotemark{a} \\*
   &  &  &  & F300W & 5100 & No & U4WODE04B\tablenotemark{a} \\*
   &  &  &  & F450W & 2210 & Marginal & U4WODE07B\tablenotemark{a} \\*
   &  &  &  & F814W & 3020 & No & U4WODE0CB\tablenotemark{a} \\*
   &  &  &  & F656N & 5000 & Yes & U4WODE0IB\tablenotemark{a} \\
LMC--RP~683\tablenotemark{b} & Possible (VLE) & 4.0 & WFPC2/WFALL & F606W & 3500 & Marginal
& U6GZOE01M\tablenotemark{a} \\*
   &  &  &  & F656N & 1400 & Yes & U4WOFY0AB\tablenotemark{a} \\
   &  &  &  & F300W & 7300 & No & U6GZOE02M\tablenotemark{a} \\
   &  &  &  & F450W & 1800 & Marginal & U4WOFZ04M\tablenotemark{a} \\
   &  &  &  & F814W & 1760 & No & U4WOFZ06M\tablenotemark{a} \\
LMC--RP~723\tablenotemark{b} & True & 4.7 & WFPC2/WFALL & F439W & 4x400 & Marginal & U4ZN0502B \\*
   &  &  &  & F555W & 4x400 & Yes & U4ZN050BR \\*
   &  &  &  & F675W & 820 & No & U4ZN0506B \\*
   &  &  &  & F814W & 1020 & No & U4ZN0508B \\
LMC--RP~764\tablenotemark{b} & True & 6.0 & WFPC2/WFALL & F606W & 500 & Yes & U8UOUO01M \\*
   &  &  &  & F606W & 500 & Yes & U8UOUQ01M \\*
   &  &  &  & F300W & 1500 & Marginal & U8UOUO02M \\*
   &  &  &  & F300W & 1500 & Marginal & U8UOUQ02M \\
LMC--RP~885\tablenotemark{b} & True & 3.4 & WFPC2/WFALL & F555W & 3700 & Yes & U2O90104B\tablenotemark{a} \\*
   &  &  &  & F814W & 3900 & Marginal & U2O90105B\tablenotemark{a} \\
LMC--RP~1375\tablenotemark{b} & Possible & 3.4 & WFPC2/WFALL & F656N & 2070 & Yes & U4KY0805R \\
LMC--RP~1443 & True & 6.0 & WFPC2/WFALL & F547M & 40 & No & U26M1202T \\
LMC--RP~1550\tablenotemark{b} & True & 7.0 & WFPC2/WFALL & F555W & 80 & Yes & U69W0102R \\
LMC--RP~1580 & True & 3.0 & WFPC2/WFALL & F547M & 800 & No & U64B0601B\tablenotemark{a} \\
LMC--RP~1759 & True & 6.0 & WFPC2/WFALL & F555W & 20 & No & U26M1B02P \\*
   &  &  &  & F450W & 40 & No & U26M1B01P \\
LMC--RP~2180 & Likely & 10.1 & ACS/WF1 & F555W & 20 & No & J8NE68REQ \\*
   &  &  &  & F814W & 20 & No & J8NE68RIQ \\
\enddata
\tablenotetext{a}{MAST high-level science product, composed of multiple exposures.}
\tablenotetext{b}{Extended emission detected in at least one filter.}
\end{deluxetable}
%
%
\begin{deluxetable}{lllcccl}
\tabletypesize{\scriptsize}
\tablecolumns{7}
\tablewidth{0pc}
\tablecaption {Coordinates, Dimensions, and Morphologies of the RP LMC Nebulae \label{Morph}}
\tablehead {
\colhead {} & \colhead {R.A.} & \colhead {Decl.} &
\colhead {Diameter} & \colhead {Object} & \colhead {Morph.} & \colhead {} \\
\colhead {Nebula} & \colhead {J(2000)} & \colhead {(J2000)} &
\colhead {(arcsec)} & \colhead {Class} & \colhead {Class} & \colhead {Notes} \\
\colhead {(1)} & \colhead {(2)} & \colhead {(3)} & \colhead {(4)} &
\colhead {(5)} & \colhead {(6)} & \colhead {(7)}
}
\startdata
LMC--RP~265  & 5 37 00.72  & $-69$ 21 29.1 & $4.2 \times 3.4$ & PN & B? & \\
LMC--RP~671  & 5 26 11.28 & $-70$ 16 05.6 & 4.78  & PN & R & CS identified \\
LMC--RP~683  & 5 36 36.07 & $-70$ 07 05.3 & $\sim3 \times 4$ & H~II reg.? & \nodata & Diffuse emission \\
LMC--RP~723  & 5 25 04.66 & $-69$ 48 33.3 & 3.2  & PN & R & CS identified \\
LMC--RP~764  & 5 26 16.35 & $-69$ 38 01.9 & 2.77 $\times$ 3.70  & PN & B & CS tentative identification \\
LMC--RP~885  & 5 24 13.62 & $-69$ 47 23.7 & 1.1  & PN & R & Attached shell is 2\farcs2; CS identified \\
LMC--RP~1375 & 5 18 15.09 & $-69$ 16 22.2 & $3.4 \times 4.8$  & PN? & E? & Diam. w.r.t. 10\% contour \\
LMC--RP~1550 & 5 08 49.70 & $-68$ 44 05.6 & 1.11 $\times$ 1.24  & PN & B & \\
\enddata
\end{deluxetable}

\clearpage


\end{document}